\documentclass{aip-cp}

\usepackage[numbers]{natbib}
\usepackage{rotating}
\usepackage{graphicx}

\usepackage{url}
\newcommand*\hess{H.E.S.S.}
\newcommand*\fermi{\textit{Fermi}}
\newcommand*\ctools{\textit{ctools}}
\newcommand*\astrisim{\textit{ASTRIsim}}

\begin{document}

% Title portion
\title{Prospects for PWNe and SNRs science with the ASTRI mini-array of pre-production small-sized telescopes of the Cherenkov Telescope Array}

\author[aff1,aff2]{A. Burtovoi\corref{cor1}}
\author[aff2]{L. Zampieri}
	%\eaddress{luca.zampieri@oapd.inaf.it}
\author[aff3]{A. Giuliani}
\author[aff4]{C. Bigongiari}
\author[aff4]{F. Di Pierro}
\author[aff4]{A. Stamerra}
\author{the ASTRI Collaboration}
	\eaddress[url]{http://www.brera.inaf.it/astri/}
\author{the CTA Consortium}
	\eaddress{see http://www.cta-observatory.org for full author \& affiliation list}

\affil[aff1]{Centre of Studies and Activities for Space (CISAS) ``G. Colombo'', University of Padova, Via Venezia 15, I-35131 Padova, Italy.}
\affil[aff2]{INAF - Astronomical Observatory of Padova, vicolo dell' Osservatorio 5, I-35122 Padova, Italy.}
\affil[aff3]{INAF - IASF Milano, via E.Bassini 15, I-20133 Milano, Italy.}
\affil[aff4]{INAF - Astronomical Observatory of Torino, strada Osservatorio 20, I-10025 Pino Torinese (TO), Italy.}

\corresp[cor1]{Corresponding author: aleksandr.burtovoi@studenti.unipd.it}

\maketitle

\begin{abstract}
%The development and construction of the Cherenkov Telescope Array (CTA), in which INAF is involved through the ASTRI project, opens up new opportunities for the study of very high energy (VHE, $E>100$ GeV) sources. 
The development and construction of the Cherenkov Telescope Array (CTA) opens up new opportunities for the study of very high energy (VHE, $E>100$ GeV) sources. As a part of CTA, the ASTRI project, led by INAF, has one of the main goals %One of the ASTRI project goals is 
to develop one of the mini-arrays of CTA pre-production telescopes, proposed to be installed at the CTA southern site. Thanks to the innovative dual-mirror optical design of its small-sized telescopes, the ASTRI mini-array will be characterized by a large field of view, an excellent angular resolutioerrorn and a good sensitivity up to energies of several tens of TeV. Pulsar wind nebulae, along with Supernova Remnants, are among the most abundant sources that will be identified and investigated, with the ultimate goal to move significantly closer to an understanding of the origin of cosmic rays (CR). As part of the ongoing effort to investigate the scientific capabilities for both CTA as a whole and the ASTRI mini-array, we performed simulations of the Vela X region. We simulated its extended VHE $\gamma$-ray emission using the results of the detailed \hess~analysis of this source. %We simulated its extended VHE $\gamma$-ray emission assuming that its morphology is intermediate between those in the radio and X-ray bands, and assuming the H.E.S.S. spectrum. 
We estimated the resolving capabilities of the diffuse emission and the detection significance of the pulsar with both CTA as a whole and the ASTRI mini-array. Moreover with these instruments it will be possible to observe the high-energy end of SNRs spectrum, searching for particles with energies near the cosmic-rays ``knee'' ($E\sim10^{15}$ eV). We simulated a set of ASTRI mini-array observations for one young and an evolved SNRs in order to test the capabilities of this instrument to discover and study PeVatrons on the Galactic plane.
\end{abstract}

\section{INTRODUCTION}
The Cherenkov Telescope Array (CTA) is ground-based facility project, which will observe the very high energy (VHE, $E>100$ GeV) $\gamma$-ray sky \citep{CTA2011}.  It will be an Imaging Atmospheric Cherenkov Telescopes (IACTs) instrument with 10 times better sensitivity and improved angular resolution (see e.g. \citep{Acharya2013}), compared to the present IACT installations such as MAGIC, VERITAS and \hess~CTA is designed as an array of a large number of telescopes of 3 different types, comprising a few Large Size Telescopes (LSTs, 23-meter diameter), several dozens of Medium Size Telescopes (MSTs with diameters of 10--12 meters) and $\sim$70 Small Size Telescopes (SSTs, 4-meter diameter). This will allow us to perform observations in a wide energy range from a few tens of GeV to more than 100 TeV. The CTA observatory will comprise two arrays, one in each hemisphere for full coverage of the sky. 

As a part of the CTA project, a dual-mirror end-to-end prototype of the Small Size Telescope (SST-2M) is under undergoing commissioning activities within the framework of the ASTRI project of the Italian Ministry of Research and Education led by the Italian National Institute of Astrophysics (INAF) \citep{LaPalombara2014}. The prototype was recently inaugurated in Italy (Mount Etna, Sicily) \citep{Vercellone2015arx}. The innovative dual-mirror Schwarzschild-Couder optical design \citep{Vassiliev2007} implemented for the ASTRI telescopes allows us to reduce the size and the weight of the Cherenkov camera and to adopt silicon photo-multipliers as light detectors. In cooperation with the Universidade de Sao Paulo (Brazil) and the North-West University (South Africa), this project also foresees the construction of a mini-array of 9 ASTRI SST-2M units with a relative distance of about 300 m (ASTRI mini-array). The ASTRI mini-array will possibly be one of the precursors for the CTA southern array. An improved sensitivity above 10 TeV with respect to the current IACTs ($\sim$2.5 times better than that of \hess~at 30 TeV), together with an angular resolution of few arc minutes and energy resolution of 10--15\% will make the ASTRI mini-array a very promising facility for observations of bright known Galactic and extragalactic objects at energies up to few hundreds of TeV. In addition, ASTRI mini-array will cover relatively large area on the sky due to the $\sim$10$^\circ$ field of view of each dual-mirror ASTRI SST-2M telescope \citep{Vercellone2013}.

Pulsar wind nebulae (PWNe) and supernova remnants (SNRs) are very interesting targets for $\gamma$-ray observations with IACTs. Such objects correspond to the most abundant classes of Galactic VHE sources. They comprise regions where particles are accelerated up to relativistic energies. CTA and the ASTRI mini-array will be able to carry out more detailed observations of these sources as compared to present generation of IACTs and will shed more light on the problem of the origin of the Galactic cosmic rays (CR). 

In this work we investigated the scientific capabilities of CTA and the ASTRI mini-array, simulating VHE observations of the Vela PWN and two SNRs -- young RCW 86 and evolved W 28. Making a number of assumptions, we examine resolving capabilities of these instruments observing extended sources such as the Vela PWN and SNR W 28, the detectability of the Vela pulsar. We also checked if the ASTRI mini-array will be able to discriminate between hadronic and leptonic nature of the $\gamma$-ray emission from SNR RCW 86.

\section{VHE SIMULATIONS}
We perform simulations using different software packages such as \ctools~(see \citep{Knodlseder2016} and \url{http://cta.irap.omp.eu/ctools/}) and \astrisim~(for more details see \url{http://www.iasf-milano.inaf.it/~giuliani/astrisim/}). CTA observations of the Vela X region are simulated with the software \ctools, whereas the ASTRI mini-array simulations of Vela X, RCW 86 and W 28 are carried out using an alternative package \astrisim, which is developed by the ASTRI Collaboration.

%\subsection{\ctools~Software}
The \ctools~simulations of the Vela X region are performed for the complete southern CTA installation (CTA-South). This array contains 4 LSTs, 24 MSTs and 72 SSTs. Instrument response functions (IRFs) of CTA-South available at \url{https://portal.cta-observatory.org/Pages/CTA-Performance.aspx} are calculated using MC-Prod2 simulations of a 50-hour observation of a point source with a flux of 1 Crab Unit ($2.79\times10^{-11} \times (E/{\rm 1~TeV})^{-2.57}$ cm$^{-2}$ s$^{-1}$ TeV$^{-1}$), located at  20$^{\circ}$ zenith angle and at the center of the field of view (see e.g. \citep{Hassan2015arx}).
We use the tool \texttt{ctobssim} to perform the simulations and obtain the event lists of the observations, whereas the tool \texttt{ctbin} -- to convert them into count maps. Residual maps are obtained using the tool \texttt{csresmap}. With this tool we subtract the background modeled with \texttt{ctmodel} from the simulated sky-map. The tool \texttt{csresmap} was used with an algorithm \texttt{SUB} (for more details see \url{http://cta.irap.omp.eu/ctools/reference_manual/csresmap.html}).
The spectral fit of the simulated data have been done with the \texttt{ctlike} tool in unbinned mode. Using \texttt{ctlike} we estimate also the significance of each source in the model, performing a maximum likelihood analysis of the simulated CTA data.

%\subsection{\astrisim~Software}
We also used an alternative software \astrisim, developed for simulating the ASTRI mini-array observations. \astrisim~is a  fast scientific software, which can simulate both point-like and extended sources. It is fully compatible with \ctools~and it makes use of IRFs computed by the full Monte Carlo code (e.g. \citep{Hassan2015arx}). It also includes spectral analysis tools (see \url{http://www.iasf-milano.inaf.it/~giuliani/astrisim/simulations/}). As a standard input, \astrisim~can use the files from the TeV-spectra catalog available at \url{http://www.iasf-milano.inaf.it/~giuliani/astrisim/gsed/}.
We performed the ASTRI mini-array simulations using the IRF of the configuration (Conf.) s9-4-257m, provided within the framework of the Sep 2014 Monte Carlo Prod2 DESY package, site ``Leoncito++'', available at \url{http://www.cta-observatory.org/ctawpcwiki/index.php/WP_MC\#Interface_to_WP_PHYS}. The Conf. s9-4-257m represents an array of 9 dual-mirror SSTs separated by 257 m.

\section{IMAGING THE VELA X REGION WITH CTA \& ASTRI MINI-ARRAY}
Vela X is a well-studied source, associated with the Vela pulsar (PSR J0835$-$4510) and to an extended relatively old pulsar wind nebula (age $\gtrsim$10$^4$ years \citep{Lyne1996}). The Vela pulsar was investigated in $\gamma$-rays with the \fermi-LAT \citep{Abdo2010_3} and recently with the \hess~telescope (\url{http://www.mpg.de/8287998/velar-pulsar}). The Vela PWN was studied at all wavelengths. Recent 53.1 h observations made by \hess~shows a complex morphology of this source at VHE from 0.75 up to 70 TeV \citep{Abramowski2012}. In contrast to previous \hess~studies \citep{Aharonian2006_3}, significant extended emission out of the X-ray cocoon (``wings'') was found. This challenged the theoretical leptonic scenario (see e.g. \citep{deJager2008}), which explains multiwavelength emission from Vela X due to the presence of two population of leptons: (i) the first population of energetic particles produces synchrotron X-ray and inverse Compton (IC) VHE photons inside the cocoon and (2) the second population of low-energy particles produces synchrotron radio and GeV $\gamma$-ray photons due to IC outside the cocoon (see e.g. \citep{deJager2008,Abdo2010_2,Grondin2013}). Some models consider the emission of accelerated hadrons ($\pi^0$-decay) as the main mechanism of the VHE emission from the cocoon (see e.g. \citep{Horns2006}). Another problem is to explain the lack of variations of VHE spectra measured by \hess~in the inner and outer parts of the Vela PWN. Future, more detailed spectral and morphological investigations of Vela X with CTA and the ASTRI mini-array will be very important for improving our understanding of this source and clarifying the nature of its broadband emission (leptonic or hadronic origin).

To simulate the VHE emission from the Vela X region, we define the spatial and spectral models of all sources in this area of the sky (see Table \ref{tab:4.1}). The Vela pulsar is simulated as a point source with a power-law spectrum obtained from the analysis of the 5 years of the \fermi-LAT data at energies $>$10 GeV and extrapolated to the VHE range. To define the spatial model of the extended Vela X emission, we created radio and X-ray templates, adopting archival high-resolution observations of MOST and ROSAT telescopes obtained during the second Molongo Galactic Plane Survey \citep{Bock1999,Murphy2007} and the ROSAT All Sky Survey \citep{Voges1999}, respectively. Following Abramowski et al. \citep{Abramowski2012}, we assumed that the VHE emission from the Vela PWN is produced by the same leptonic populations, which are responsible for the X-ray cocoon emission and radio extended ``wings'', and that the Vela PWN morphology can be approximated by the superposition of the radio (65\% in flux) and X-ray (35\% in flux) spatial maps. We then select circular and ring regions in both the X-ray and radio templates (for more details see Table \ref{tab:4.1} and \citep{Abramowski2012}) and perform analysis of the Vela PWN simulating circular and ring regions of our templates with different power-law with exponential cut-off models (PLEC), which were obtained by \hess~in the 0.75-70 TeV energy range \citep{Abramowski2012}. 
%Following Abramowski et al. \citep{Abramowski2012}, we assumed that the VHE emission from the Vela PWN is produced by the same leptonic populations, which are responsible for the X-ray cocoon emission and radio extended ``wings'', and that the Vela PWN morphology can be approximated by the superposition of the radio (65\% in flux) and X-ray (35\% in flux) spatial maps. For the spatial models of the extended Vela PWN emission we created radio and X-ray templates (circular and ring regions, for more details see \citep{Abramowski2012}), adopting archival high-resolution observations of MOST and ROSAT telescopes obtained during the second Molongo Galactic Plane Survey \citep{Bock1999,Murphy2007} and ROSAT All Sky Survey \citep{Voges1999}, respectively. For the VHE spectrum of the Vela PWN we used the  power-law models with exponential cut-off (PLEC) from the analysis of the \hess~data in the 0.75-70 TeV energy range \citep{Abramowski2012}. 
Parameters of all components contributing to the VHE $\gamma$-ray spectrum from the Vela X region are listed in Table \ref{tab:4.1}.

\begin{table*}
\footnotesize
\caption{Parameters of the components contributing to the VHE $\gamma$-ray spectrum from the Vela X region.}
\label{tab:4.1}
\centering
\tabcolsep7pt
\begin{tabular}{l c c l}
\hline
\tch{1}{c}{b}{Component}	& \tch{1}{c}{b}{Spatial model}	& \tch{1}{c}{b}{Type of the spectrum}	& \tch{1}{c}{b}{Spectral parameters}	\\
\hline
PSR J0835$-$4510	& point source:	& power law	& $N_{\rm PSR}=19.4 \times 10^{-14}$ cm$^{-2}$ s$^{-1}$ MeV$^{-1}$	\\ 
			& $\alpha_*=128.83^{\circ}$,	&	& $\gamma=4.45$		\\
			& $\delta_*=-45.18^{\circ}$	& 	& $E_0 = 20$ GeV		\\
\hline
Vela X 		& diffuse source:	& PLEC		& $N_{\rm X1} = 11.6 \times 10^{-12}$ cm$^{-2}$ s$^{-1}$ TeV$^{-1}$	\\
circular region	 &65\% of radio + & 				& $\gamma = 1.36$		\\
 ($<$0.8$^{\circ}$) &35\% of X-ray &			& $E_c = 13.9$ TeV		\\
			&templates & 			& $E_0 = 1$ TeV		\\
%&&&\\
\hline
Vela X 		& diffuse source:	& PLEC		& $N_{\rm X2} = 3.3 \times 10^{-12}$ cm$^{-2}$ s$^{-1}$ TeV$^{-1}$		\\
ring			&65\% of radio + &		& $\gamma = 1.14$		\\
($0.8^{\circ}$-$1.2^{\circ}$)&35\% of X-ray &			& $E_c = 9.5$ TeV		\\
			&templates & 			& $E_0 = 1$ TeV		\\
\hline
Isotropic 		& Gaussian radial	& \multicolumn{2}{c}{dominated by the cosmic-ray}\\
background	& profile			& \multicolumn{2}{c}{background}			\\
\hline
\multicolumn{4}{p{0.88\textwidth}}{$N_{\rm PSR}$, $N_{\rm X1}$ and $N_{\rm X2}$ are the normalization factors. $\gamma$ is the spectral index. $E_c$ and $E_0$ are the cut-off and scale energies, respectively.}
\end{tabular}
\end{table*}

We obtained VHE residual maps of Vela X using simulations performed with \ctools~and \astrisim. The 50 h observations of the Vela X region with the CTA-South array and the ASTRI mini-array are shown in Figures \ref{fig:VelaX_Maps_ROSSAT_MOST} and \ref{fig:VelaX_ASTRIsim_R_M_E}, respectively. Since the resolving capability of a typical IACT improves with energy, the emission from the Vela X region was investigated in different energy ranges: $E$$>$0.04 TeV, $>$0.25 TeV and $>$1 TeV for CTA-South and $E$$>$1.6, $>$10 and $>$50 TeV for the ASTRI mini-array. In Figures \ref{fig:VelaX_Maps_ROSSAT_MOST} and \ref{fig:VelaX_ASTRIsim_R_M_E} we also smoothed residual maps according to the different angular resolutions of the corresponding instruments. Smoothing radii adopted here are equal to the 68\% of the containment radii ($r_{68}$) of the $\gamma$-ray point spread function (PSF) at the lower limit of the corresponding energy ranges. For the CTA-South $r_{68} = 0.19^{\circ}$ at 0.04 TeV, $r_{68} = 0.09^{\circ}$ at 0.25 TeV and $r_{68} = 0.05^{\circ}$ at 1 TeV. For the ASTRI mini-array $r_{68} = 0.14^{\circ}$ at 1.6 TeV, $r_{68} = 0.06^{\circ}$ at 10 TeV and $r_{68} = 0.08^{\circ}$ at 50 TeV.

\newcommand*\XMw{0.22}
\begin{figure*}[t]
	\includegraphics[height=\XMw\textheight]{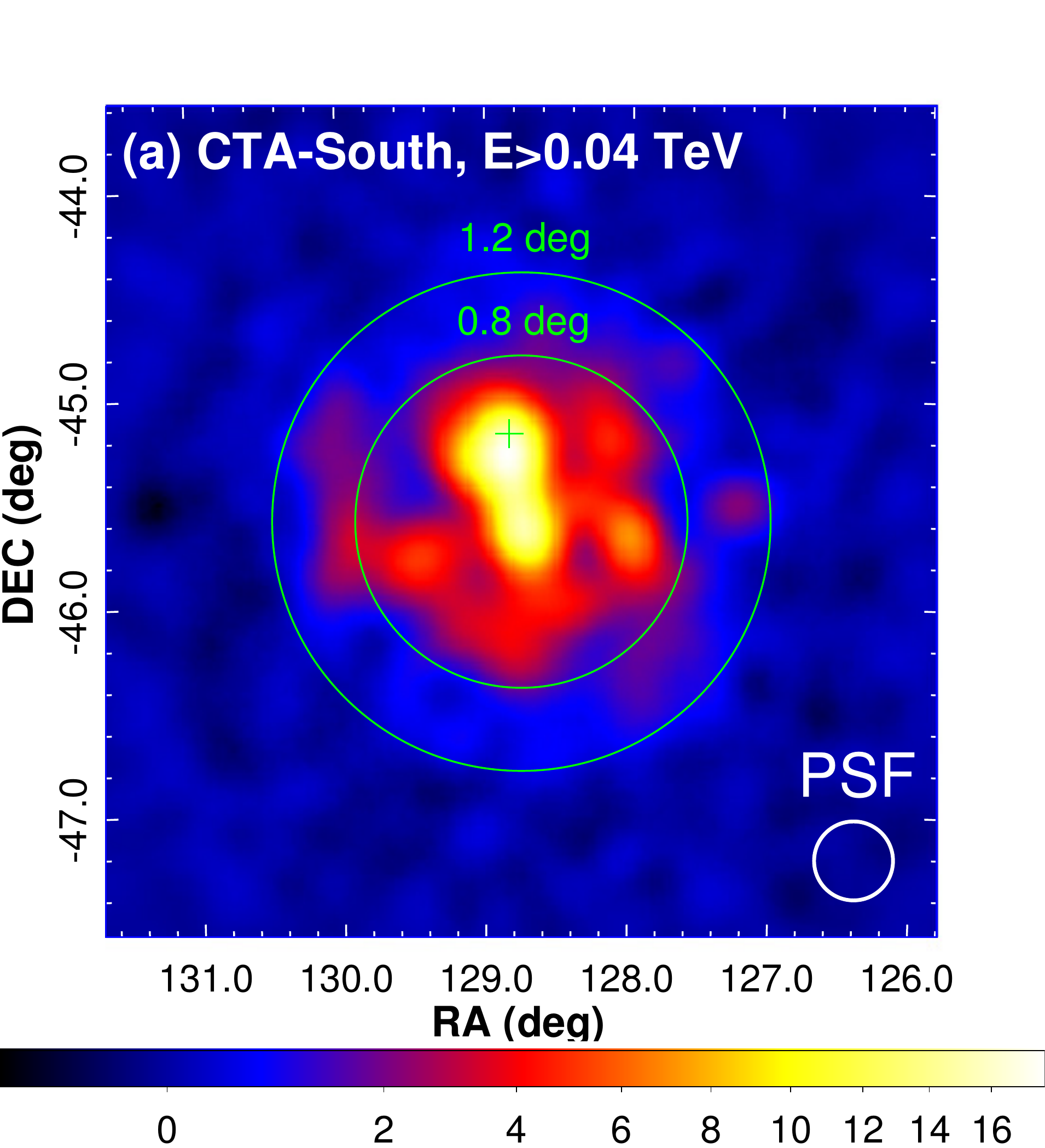}
	\hfill
	\includegraphics[height=\XMw\textheight]{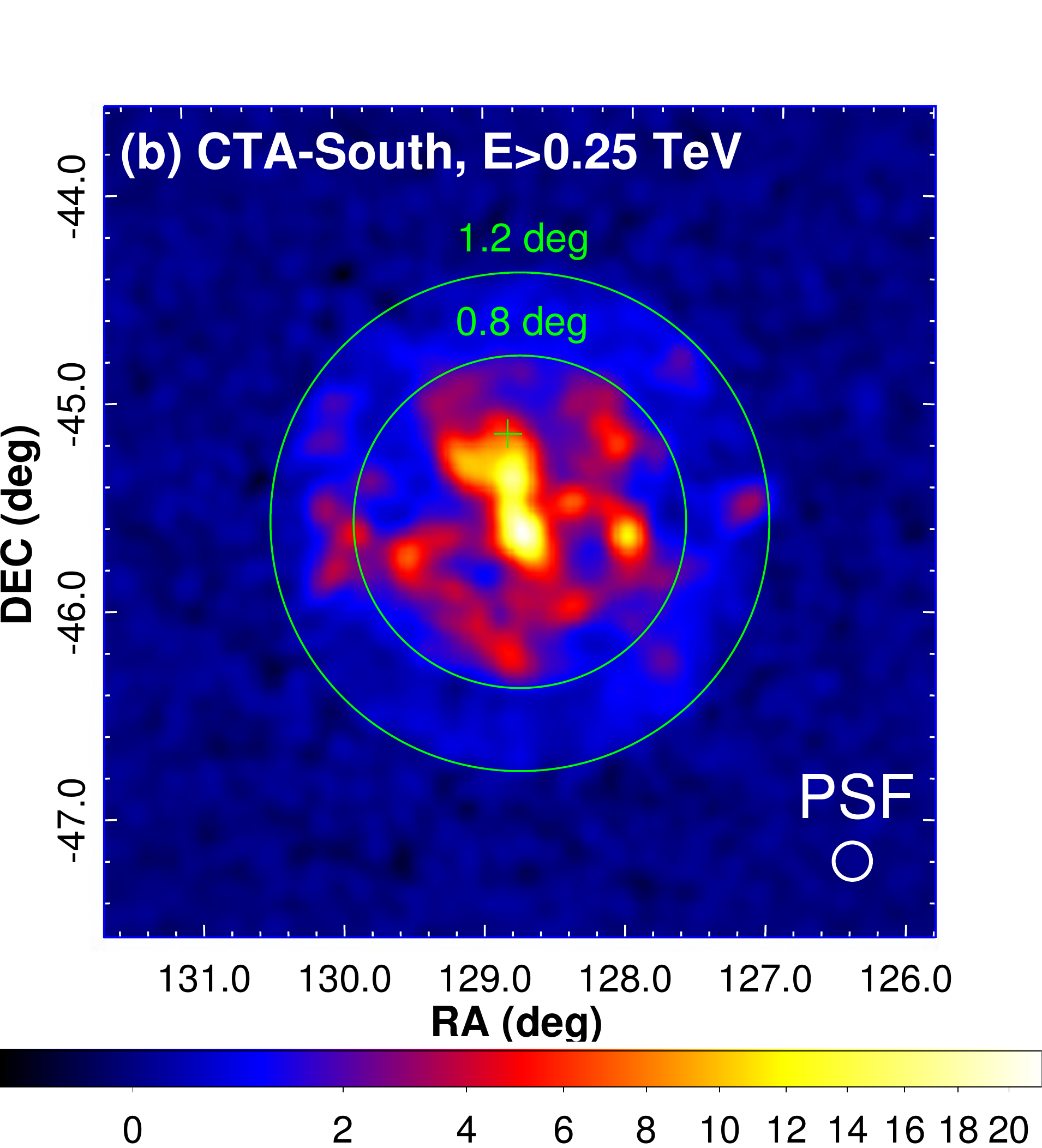}
	\hfill
	\includegraphics[height=\XMw\textheight]{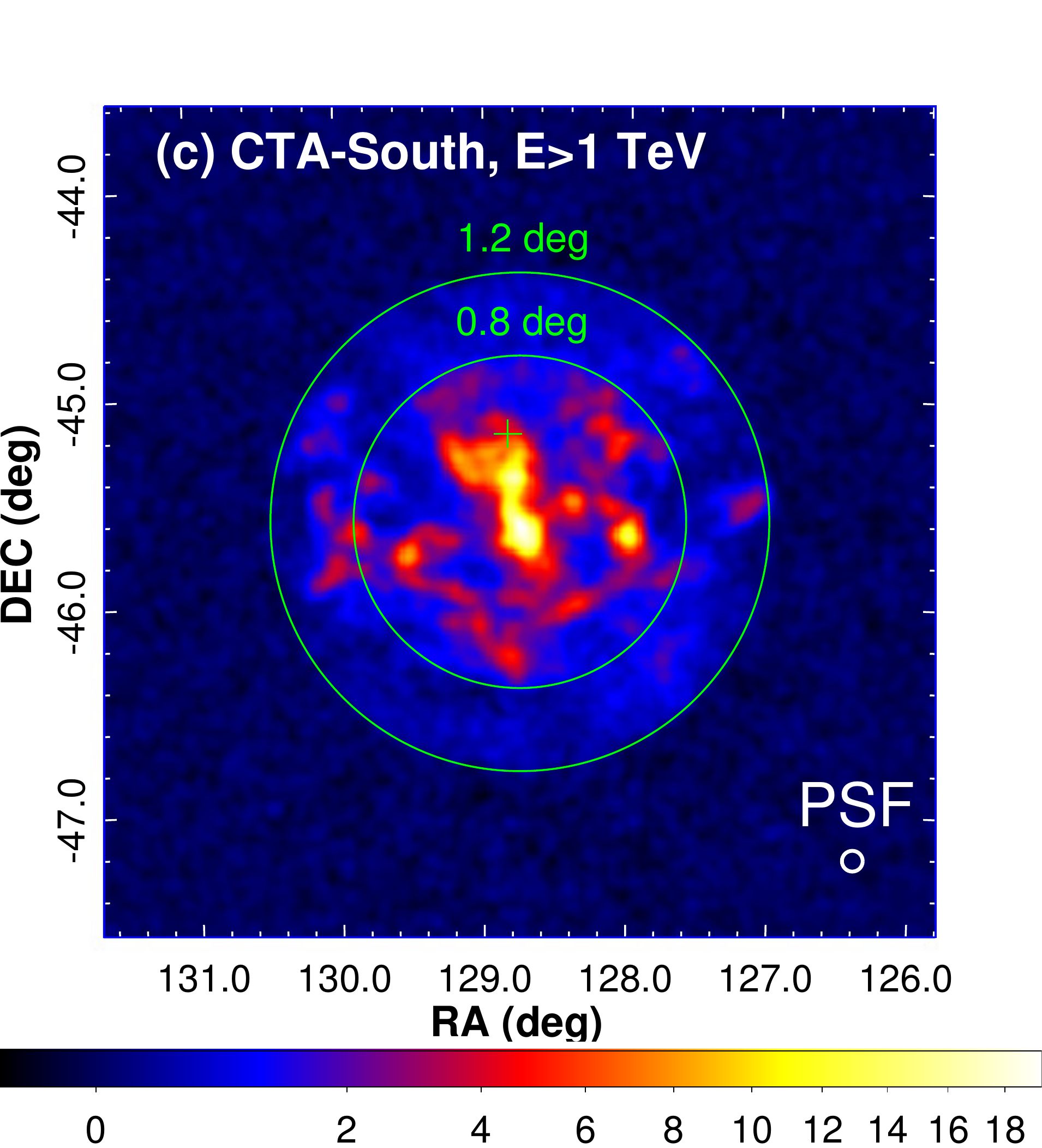}
	\caption{VHE Vela X residual maps simulated with \ctools~assuming 65\%-contribution from the MOST radio template and 35\%-contribution from the X-ray template. The CTA-South IRF is used. The exposure time is 50 hours. Energy ranges are (a) $E>0.04$ TeV; (b) $E>0.25$ TeV; (c) $E>1$ TeV. $x$-, $y$- axes are the right ascension (RA) and declination (DEC) in degrees. The $4^{\circ}\times4^{\circ}$ field of view is centered at $\alpha_0=128.75^{\circ}$ and $\delta_0=-45.6^{\circ}$. The color bars represent residual number of counts  in the square root scale. All maps are smoothed according to the size of the corresponding PSF (white circles). See text for details. Green circles have radii of 0.8$^{\circ}$ and 1.2$^{\circ}$, respectively. The green cross marks the Vela pulsar position.}
	\label{fig:VelaX_Maps_ROSSAT_MOST}
\end{figure*}

\newcommand*\Aw{0.22}
\begin{figure*}[t]
	\includegraphics[height=\Aw\textheight]{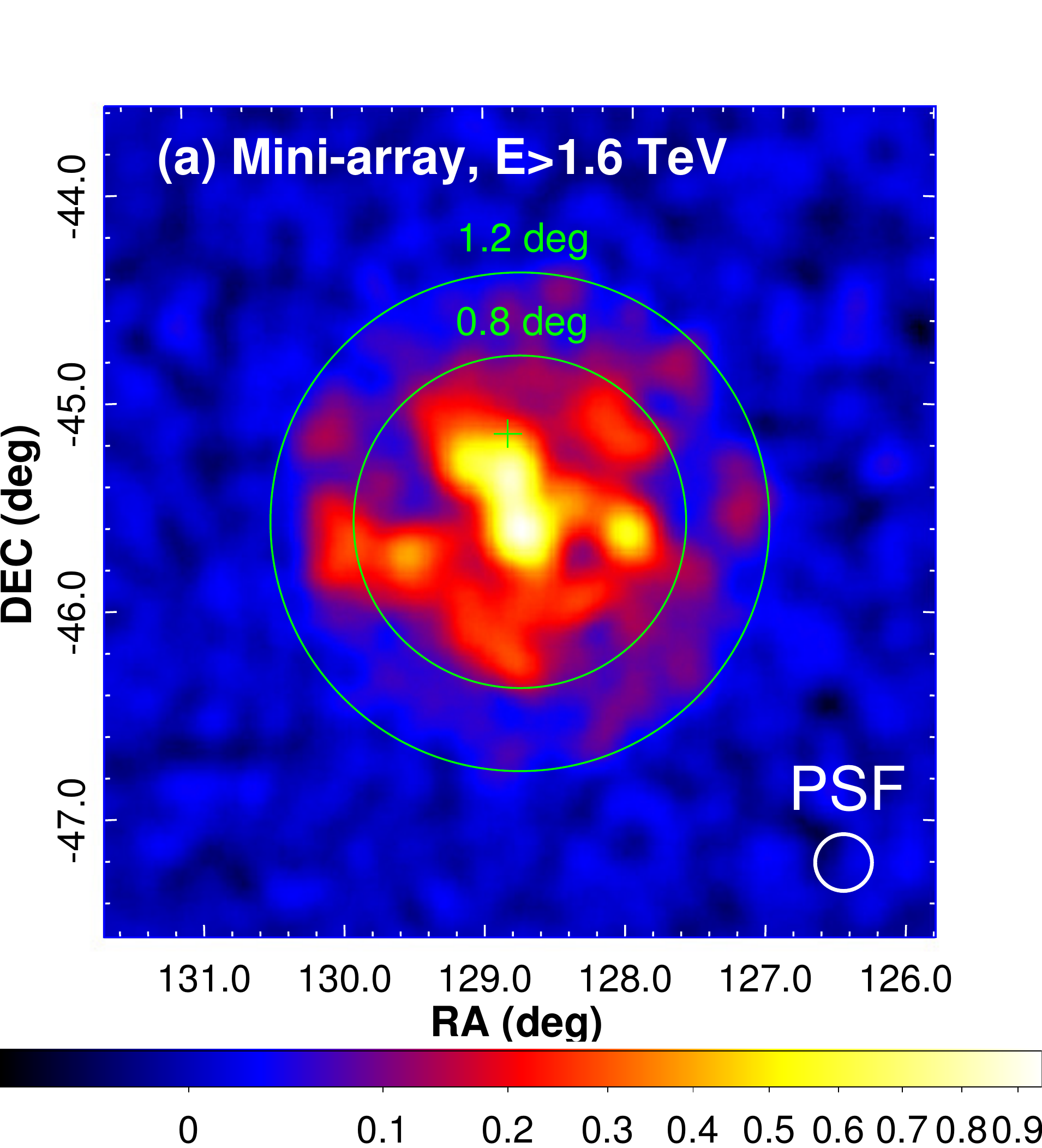}
	\hfill
	\includegraphics[height=\Aw\textheight]{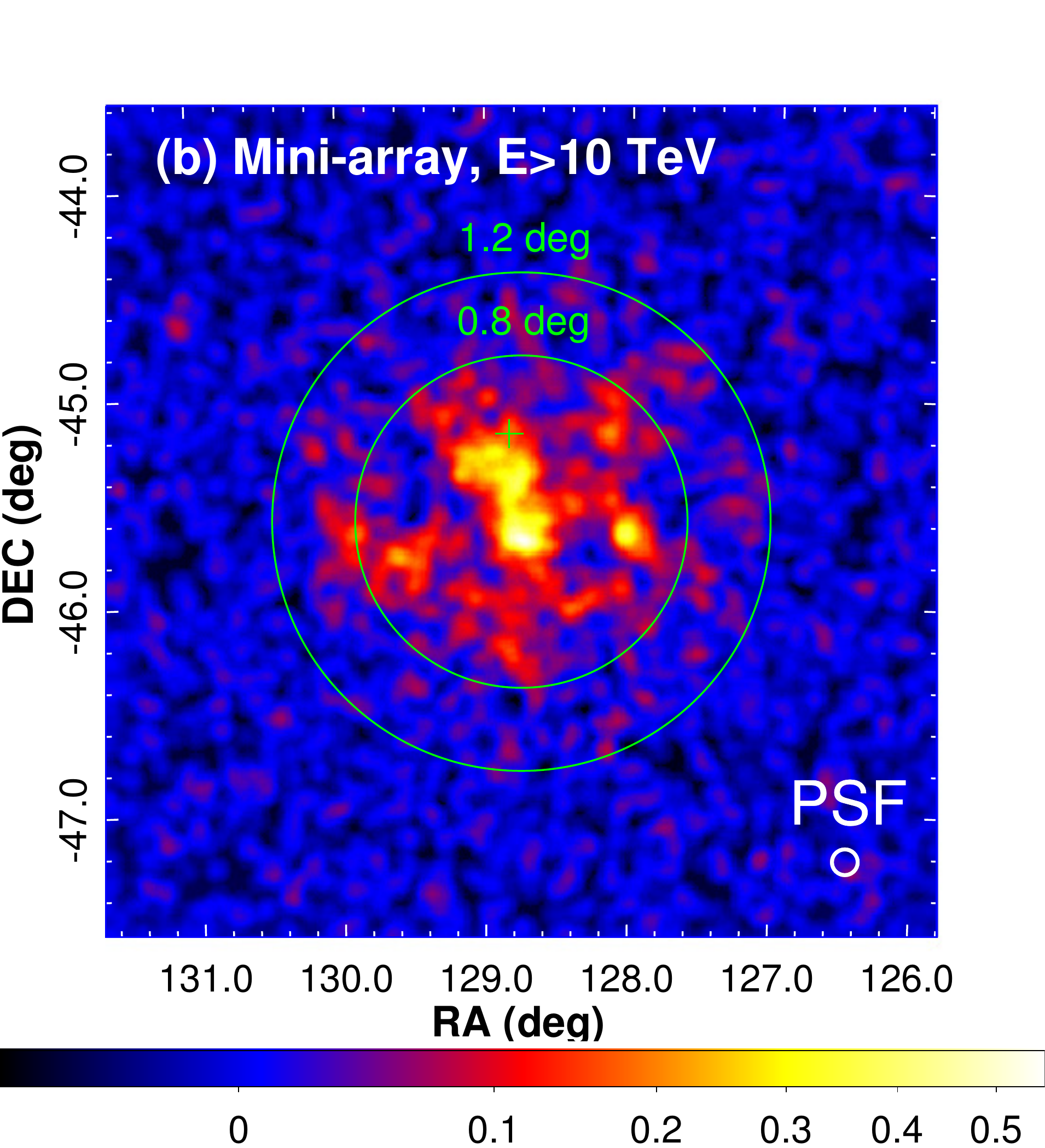}
	\hfill
	\includegraphics[height=\Aw\textheight]{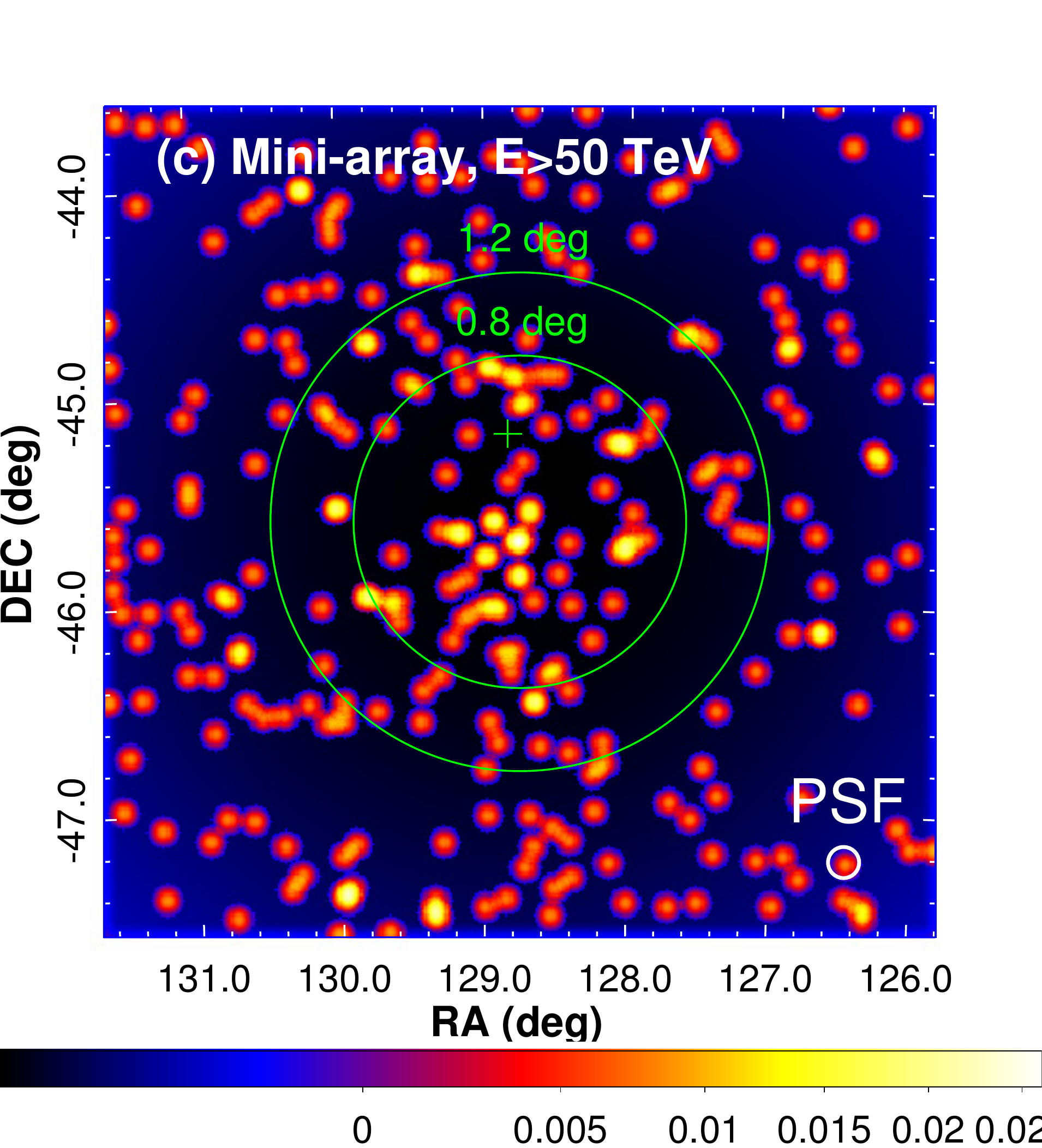}
	\caption{Same as Figure \ref{fig:VelaX_Maps_ROSSAT_MOST} for the ASTRI mini-array (Conf. s9-4-257m, Mini-array) and for the energy ranges (a) $E>1.6$ TeV; (b) $E>10$ TeV; (c) $E>50$ TeV, simulated with \astrisim.}
	\label{fig:VelaX_ASTRIsim_R_M_E}
\end{figure*}

The resulting significance $S$ of the Vela pulsar detection with CTA-South during a 50-hour exposure is the following: (i) $S=39.1$ for $E>0.04$ TeV, (ii) $S=13.9$ for $E>0.1$ TeV and (iii) $S=4.5$ for $E>0.25$ TeV. For higher energies unrealistically long duration of observation is required for the significant Vela pulsar detection.

%\begin{table*}
%\caption{Significance $S$ of the Vela pulsar detection with CTA-South and the ASTRI mini-array (Conf. s9-4-257m, Mini-array) in different energy ranges. Duration of observation is 50 hours.}
%\label{tab:4.2}
%\centering
%\tabcolsep7pt
%\begin{tabular}{l l c l}
%\hline
%\tch{1}{c}{b}{Array}	& \tch{1}{c}{b}{Energy range}	& \tch{1}{c}{b}{Significance $S$}	\\
%\hline
%CTA-South: 	&&	\\
%			& $E>0.04$ TeV	& 39.1		\\
%			& $E>0.1$ TeV		& 13.9		\\
%			& $E>0.25$ TeV	& 4.5		\\
%			& $E>0.75$ TeV	& not signif.	\\
%			& $E>1$ TeV		& not signif.	\\
%Mini-array: 	&&	\\
%			& $E>1.6$ TeV		& not signif.	\\
%\hline
%\end{tabular}
%\end{table*}

%...Discussion...
The Vela PWN will be detected (5$\sigma$) in about 10 min with CTA and in a few hours with the ASTRI mini-array. If our assumption that the extended TeV emission from Vela X follows the radio filaments- and arcs- structure seen by MOST is the right one, the quality of the CTA data will be such that high-resolution observations can confirm it. CTA will help us to determine the contributions of the radio and X-ray populations to the VHE gamma-ray morphology with an accuracy of a few percent. We also calculated significances of the radio and X-ray components, assuming different contribution of corresponding populations of leptons. We obtained that these two components can be distinguished with CTA, if the contribution from either the radio or X-ray population is more than 10\% of the total VHE flux of Vela X. Such detailed observations will improve our understanding of the nature of the Vela X extended emission. In addition, CTA will significantly detect the Vela pulsar at VHE in a $\sim$50 h observation.

\section{PEVATRON SEARCH WITH THE ASTRI MINI-ARRAY} 
We simulated the ASTRI mini-array observations of different SNRs, such as RCW 86 and W28, 
%young SNR RCW 86 and evolved SNR W 28, 
testing the capabilities of this instrument to discover Galactic PeVatrons.

%\subsection{Young SNR: RCW 86}
RCW 86 is a fairly young shell-type SNR ($\sim$2000 yr), observed at all wavelengths. Non-thermal emission from this SNR was detected in the radio and X-rays bands. The gamma-rays emission of RCW 86 was detected by the \fermi~and \hess~telescopes. Despite of a number of detailed investigations at GeV and TeV bands, the origin (hadronic or leptonic) of $\gamma$-rays from this source is still under debate (for the most recent $\gamma$-ray studies see \citep{Aharonian2009,Yuan2014,Ajello2016,Abramowski2016arx}). As shown in figure \ref{fig:rcw86}, the obtained $\gamma$-ray data points are compatible with both the leptonic and hadronic models.
%- Debated origin: Is it interacting with molecular clouds (MCs) or it is RX J1713-like source?
The simulated observations show that the ASTRI mini-array can extend the measure of the spectrum up to 40 or 60 TeV, depending on the emission mechanism. The shape of the spectrum at such energies can distinguish both the two models and measure the maximum energy of the particles currently accelerated by SNR RCW 86. At this scope we simulated 200 h observations assuming leptonic and hadronic spectra for RCW 86. 
Results are shown in Figure \ref{fig:rcw86}. The leptonic model deviates from hadronic simulated data by 5.7$\sigma$ and vice-versa -- by 4.2$\sigma$. We obtained that the ASTRI mini-array will be able to discriminate between these two scenarios and (if hadronic) it can check the presence of PeV cosmic rays ($E_p\sim10^{15}$ eV).

\newcommand*\rcw{0.23}
\begin{figure}[t]
	\includegraphics[height=\rcw\textheight]{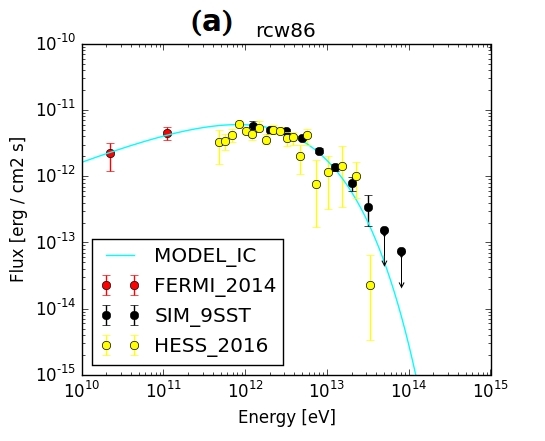}
	\hfill
	\includegraphics[height=\rcw\textheight]{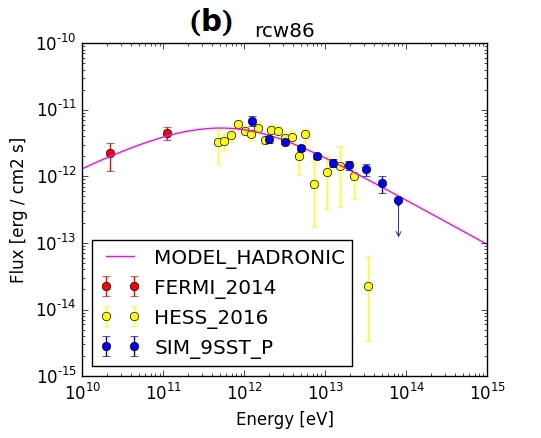}
	\caption{Simulated spectra of the SNR RCW 86 observed with the ASTRI mini-array in case if (a) leptonic (cyan line) or (b) hadronic (magenta line) mechanism of the $\gamma$-ray origin dominates. ASTRI simulated data points are in the (a) black and (b) blue colors.  \hess~observational data points are in yellow (Hess\_2016, from \citep{Abramowski2016arx}). \fermi-LAT data shown as red points (Fermi\_2014, from \citep{Yuan2014}).}
	\label{fig:rcw86}
\end{figure}

W 28 is an evolved SNR (20 000--30 000 yrs), which is interacting with a giant molecular cloud (MC) system of about $10^5$ solar masses. These MCs shine in $\gamma$-rays from 100 MeV to $\sim$10 TeV as observed by AGILE, \fermi~and \hess~\citep{w28_agile,w28_fermi,Aharonian2008}. The spectra above 10 GeV of sources HESS J1801$-$233 and HESS J1800$-$240 (see Fig. \ref{fig:w28}) are compatible with a power law of index 2.7. The modelization of this system involves CRs accelerated in the past by the SNR and still present in the ambient due to a very slow diffusion of charged particles in this site. The observation of this object in the range 10--100 TeV with good sensitivity can investigate the presence of CR with energy up to 1 PeV, testing the hypothesis that this SNR was a Pevatron in the past.
We simulated the 200 h observation of W 28 with the ASTRI mini-array. Spectrum and morphology are taken from \hess~observations \citep{Aharonian2008}. We assumed that the spots seen by \hess~are made by point-like sources. In Figure \ref{fig:w28} we compare resulting sky-map with that obtained with \hess~\citep{Aharonian2008}.
With better sensitivity, the ASTRI mini-array can continue investigation of this object at very high energies ($E>10$ TeV), studying the diffusion of CR far from the SNR shell (blue circle in Figure \ref{fig:w28}).

\newcommand*\w{0.23}
\begin{figure}
	\includegraphics[height=\w\textheight]{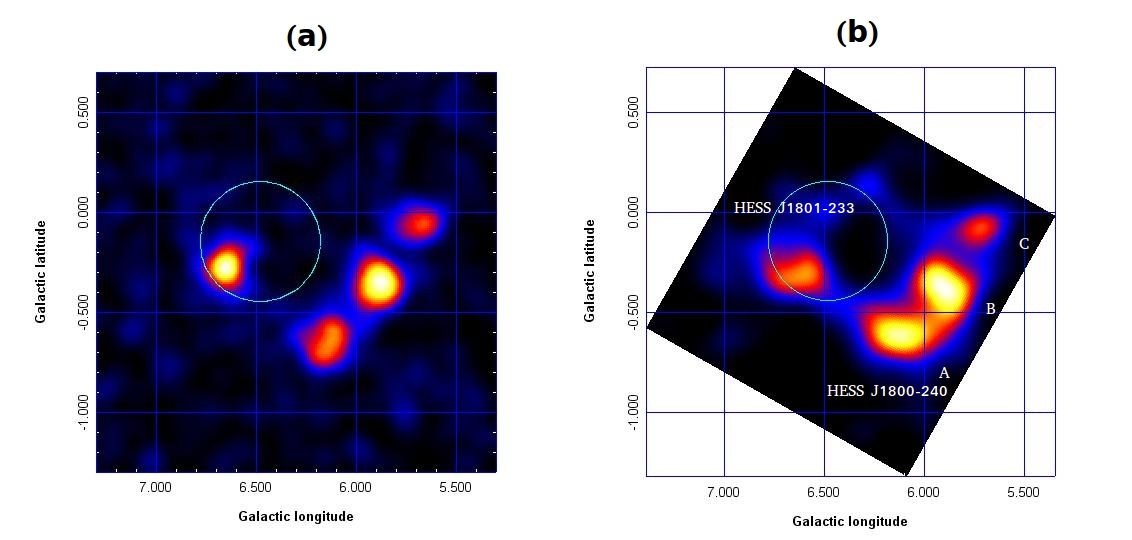}
	\caption{Sky-maps of the SNR W 28. (a) Simulated 200 h ASTRI mini-array observations, (b) \hess~40 h observations from \citep{Aharonian2008}. The blue circle represents the SNR seen in radio.}
	\label{fig:w28}
\end{figure}

\section{CONCLUSIONS}
CTA will be a very powerful instrument for the investigation of sources such as PWNe and SNRs, which accelerate particles to the relativistic energies inside our Galaxy. Better capabilities of CTA as compare to current IACTs will allow us to perform more detailed spectral and morphological analysis of VHE sources, which can lead to the understanding the origin of the Galactic CR.

Our simulations show that very promising early studies will be possible to carry out $>$10 TeV with the mini-array of several ASTRI units. With unprecedented flux sensitivity of the ASTRI mini-array, observations of the Vela X region can shed more light on the nature of its extended TeV emission. Measuring the spectrum of the SNR RCW 86, we will discriminate between hadronic and leptonic scenarios. Finally,  observing SNR W 28 with the mini-array we can study the diffusion of CR far from the SNR shell.

% Acknowledgement
\section{ACKNOWLEDGMENTS}
This research made use of \ctools, a community-developed analysis package for IACT data. \ctools is based on GammaLib, a community-developed toolbox for the high-level analysis of astronomical $\gamma$-ray data. This work is supported by the Italian Ministry of Education, University, and Research (MIUR) with funds specifically assigned to the Italian National Institute of Astrophysics (INAF) for the Cherenkov Telescope Array (CTA), and by the Italian Ministry of Economic Development (MISE) within the “Astronomia Industriale” program. We acknowledge support from the Brazilian Funding Agency FAPESP (Grant 2013/10559-5) and from the South African Department of Science and Technology through Funding Agreement 0227/2014 for the South African Gamma-Ray Astronomy Programme. We also gratefully acknowledge support from the agencies and organizations listed under Funding Agencies at http://www.cta-observatory.org/ and from CISAS-University of Padova under the project ``Technological development and scientific exploitation of Aqueye+ and Iqueye for High Time Resolution Astronomy''.

% References
\newcommand*\aap{A\&A}
\let\astap=\aap
\newcommand*\aapr{A\&A~Rev.}
\newcommand*\aaps{A\&AS}
\newcommand*\actaa{Acta Astron.}
\newcommand*\aj{AJ}
\newcommand*\ao{Appl.~Opt.}
\let\applopt\ao
\newcommand*\apj{ApJ}
\newcommand*\apjl{ApJ}
\let\apjlett\apjl
\newcommand*\apjs{ApJS}
\let\apjsupp\apjs
\newcommand*\aplett{Astrophys.~Lett.}
\newcommand*\apspr{Astrophys.~Space~Phys.~Res.}
\newcommand*\apss{Ap\&SS}
\newcommand*\araa{ARA\&A}
\newcommand*\azh{AZh}
\newcommand*\baas{BAAS}
\newcommand*\bac{Bull. astr. Inst. Czechosl.}
\newcommand*\bain{Bull.~Astron.~Inst.~Netherlands}
\newcommand*\caa{Chinese Astron. Astrophys.}
\newcommand*\cjaa{Chinese J. Astron. Astrophys.}
\newcommand*\fcp{Fund.~Cosmic~Phys.}
\newcommand*\gca{Geochim.~Cosmochim.~Acta}
\newcommand*\grl{Geophys.~Res.~Lett.}
\newcommand*\iaucirc{IAU~Circ.}
\newcommand*\icarus{Icarus}
\newcommand*\jcap{J. Cosmology Astropart. Phys.}
\newcommand*\jcp{J.~Chem.~Phys.}
\newcommand*\jgr{J.~Geophys.~Res.}
\newcommand*\jqsrt{J.~Quant.~Spec.~Radiat.~Transf.}
\newcommand*\jrasc{JRASC}
\newcommand*\memras{MmRAS}
\newcommand*\memsai{Mem.~Soc.~Astron.~Italiana}
\newcommand*\mnras{MNRAS}
\newcommand*\na{New A}
\newcommand*\nar{New A Rev.}
\newcommand*\nat{Nature}
\newcommand*\nphysa{Nucl.~Phys.~A}
\newcommand*\pasa{PASA}
\newcommand*\pasj{PASJ}
\newcommand*\pasp{PASP}
\newcommand*\physrep{Phys.~Rep.}
\newcommand*\physscr{Phys.~Scr}
\newcommand*\planss{Planet.~Space~Sci.}
\newcommand*\pra{Phys.~Rev.~A}
\newcommand*\prb{Phys.~Rev.~B}
\newcommand*\prc{Phys.~Rev.~C}
\newcommand*\prd{Phys.~Rev.~D}
\newcommand*\pre{Phys.~Rev.~E}
\newcommand*\prl{Phys.~Rev.~Lett.}
\newcommand*\procspie{Proc.~SPIE}
\newcommand*\qjras{QJRAS}
\newcommand*\rmxaa{Rev. Mexicana Astron. Astrofis.}
\newcommand*\skytel{S\&T}
\newcommand*\solphys{Sol.~Phys.}
\newcommand*\sovast{Soviet~Ast.}
\newcommand*\ssr{Space~Sci.~Rev.}
\newcommand*\zap{ZAp}

\bibliographystyle{aipnum-cp}
%\bibliography{Gamma2016_BibTeX.bib}
%Control: year  (1) truncated

\end{document}